\documentclass[12pt]{article}
\input epsf.sty
\newcommand{\half}{{1\over 2}}

\newcommand{\be}{\begin{equation}}
\newcommand{\ee}{\end{equation}}
\newcommand{\ben}{\begin{eqnarray}\displaystyle}
\newcommand{\een}{\end{eqnarray}}
\newcommand{\refb}[1]{(\ref{#1})}
\newcommand{\p}{\partial}

\begin{document}

{}~

\def\sqr#1#2{{\vbox{\hrule height.#2pt\hbox{\vrule width
.#2pt height#1pt \kern#1pt\vrule width.#2pt}\hrule height.#2pt}}}
\def\Box{\mathchoice\sqr64\sqr64\sqr{4.2}3\sqr33}
\def\t{{\theta}}
\def\s{{\sigma}}
\def\e{{\epsilon}}
\def\half{{\textstyle{1\over 2}}}
\def\hhalf{{\scriptstyle{1\over 2}}}
\def\N{{\nabla}}
\def\Nb{{\bar\nabla}}
\def\p{{\partial}}
\def\pb{{\bar\partial}}

\centerline{
\bf Strings and Noncommutativity}
\bigskip
\bigskip
\centerline{Louise Dolan}
\bigskip
\centerline{\it Department of Physics}
\centerline{\it
University of North Carolina, Chapel Hill, NC 27599-3255}
\bigskip
\smallskip
\centerline{Chiara R. Nappi}
\smallskip
\centerline{\it Department of Physics }
\centerline{\it Princeton University, Princeton, NJ 08544}

\bigskip

\noindent

\section{Abstract}
In these talks we review some of the recent results on open
strings and noncommutative gauge theories,
starting from the early calculations of open strings
in a constant electromagnetic background.
We discuss both the  neutral string and the charged string.
In the latter case,
the scaling limit that leads to noncommutative
abelian gauge theory can be generalized to a scaling limit in which multiple
noncommutativity parameters enter.
Our approach corresponds to expanding a theory with $U(N)$ Chan-Paton
factors around a background $U(1)^N$ gauge field with
different magnetic fields in each $U(1)$.
This scaling limit can be interpreted in terms of a matrix model.
We also describe an open string model with a time-dependent
noncommutativity
parameter. This model is the open string version of a WZW model based
on a non-semi-simple group.
It has a  time-dependent background, and  a spacetime metric of the
plane wave type supported by a Neveu-Schwarz two-form potential.

\section{Introduction}

There are many  ways how string theory seems to challenge
our current understanding  of spacetime. Probably,
string dualities are the most striking example of this statement,
not to mention supersymmetry and extra dimensions. More recently, a new
fact has emerged,  the noncommutativity of spacetime variables
at the string scale, a fact that obviously
might have important implications for the structure of spacetime.
Noncommutativity originally emerged in the context of open strings,
starting from the treatment of open string field theory in \cite{Witten}.
More recently, it reappeared in the context of matrix theory compactified
on a torus \cite{connes,nek}. Finally, it showed up in the
low energy description of strings in an electromagnetic background
\cite{sw, acny}. In these notes we will concentrate in the last
approach.
Our interest is to investigate the emergence of noncommutativity in various
open string models, and compute the noncommutativity parameters there.
The question behind our investigation is to see what novel noncommutative
space geometries can emerge from string theory. We will go from the single
noncommutativity parameter of the neutral string to the multiple
noncommutative parameters of the open string in a $U(N)$ background, and
finally to the time-dependent
noncommutativity parameters in a model with a time-dependent background.

A very interesting question of course is what is the
 generalization of the Moyal-Weyl product corresponding to more general
noncommutativity situations.
   A related question is what is the
corresponding Born-Infeld action. In the $U(1)$
case the noncommutativity can be seen to emerge directly from a
rewriting/reinterpretation
of the Born-Infeld action.
How does that work in the non-abelian case? Is there an equivalent in the
time-dependent case? These are still open questions,
and we will not have much to say about them.

\section{Neutral String}

The propagator for the open {\it neutral} string in
a slowly varying background $U(1)$
gauge field was computed in \cite{acny} by solving
the equations of motion

\be
\label{emo}{\Box \,< x^i(z,\bar z) \, x^j(\zeta,\bar \zeta) > \,
= - 2\pi\alpha' \delta^2(z-\zeta)\, G^{ij}}
\ee
with the boundary conditions

\be
\label{bcnp}{(\partial_z -\partial_{\bar z})
< x^i(z,\bar z) \, x^j(\zeta,\bar \zeta) > \,
+  2\pi\alpha' \,g^{ik} B_{k\ell} (\partial_z +\partial_{\bar z})\,
< x^\ell (z,\bar z) \, x^j(\zeta,\bar \zeta) >\,|_{z=\bar z} \, = 0}
\ee
where  $\Box \equiv 4\partial_z\partial_{\bar z}$.
Here the open string worldsheet $\Sigma$ will denote the disc
with Euclidean metric $\gamma^{\alpha\beta} = \delta^{\alpha\beta}$,
and the complex coordinates $z, \bar z$ are related to the
original strip coordinates $\sigma,\tau$ with $\tau$ rotated to
$t\equiv i\tau$ and
$-\infty\le t\le\infty \,, 0\le \sigma \le \pi\,,\,$
by $z\equiv e^{t + i\sigma}\,, \bar z\equiv e^{t - i\sigma}$
with ${\rm Im} z \ge 0$.

The propagator was found to be

\ben
\label{anprop}
&& <x^i(z,\bar z) \, x^j(\zeta,\bar \zeta) > =
- \alpha' \,[\half g^{ij}  \ln (z - \zeta)
+ \half g^{ij}  \ln (\bar z - \bar\zeta) \nonumber \\
&&\hskip30pt + ( -\half g^{ij} + G^{ij} + {\theta^{ij}\over
{2\pi\alpha'}} ) \ln (z - \zeta) \nonumber \\
&&\hskip30pt + ( -\half g^{ij} + G^{ij} - {\theta^{ij}\over
{2\pi\alpha'}} ) \ln (\bar z - \bar\zeta)\,\,
-{i\over{2\alpha'}} \theta^{ij}\,]
\een
where
\ben
\label{geeig}
G^{ij} & = & [ (g + 2\pi\alpha B)^{-1}\, g \,
(g -  2\pi\alpha B)^{-1}]^{ij} \nonumber \\
G_{ij}  &\equiv&  g_{ij} - (2\pi\alpha')^2 ( B g^{-1} B)_{ij}
=  [ (g - 2\pi\alpha B)\, g^{-1} \,
(g + 2\pi\alpha B)]_{ij} \nonumber \\
\theta^{ij} &\equiv& - (2\pi\alpha')^2 [ (g + 2\pi\alpha B)^{-1}\, B \,
(g -  2\pi\alpha B)^{-1}]^{ij}\,.
\een

In \cite{sw} the above propagator was used to compute the ``equal time''
commutator
of the string operators on the boundary
via a short distance expansion procedure \cite{sch},
and to define a noncommutativity parameter

\be
\label{nc}
[X^i,X^j]= i\theta^{ij},
\ee
where $\theta^{ij}$ is given in \refb{geeig}. This shows that the $X^i$ are
coordinates in a noncommutative space with noncommutativity parameter
$\theta$.
Indeed in \cite{sw} it was also shown that, in the scaling limit $\alpha'=
\sqrt{\epsilon}\rightarrow 0$ and $g_{ij}=\epsilon\rightarrow 0$,  a
noncommutative gauge theory emerged with field strength

\be
\label{strength}{{\hat F}_{ij} = \partial_i {\hat A}_j -\partial_j
{\hat A}_i
- i {\hat A}_i\star {\hat A}_j + i {\hat A}_j\star {\hat A}_i }
\ee
where ${\hat A}_i$ is related to an ordinary gauge potential
$A_i$ via the relation
\be
\label{map}{{ \hat F} = {1\over{1+ F \theta }} F \,,}
\ee
known as the Seiberg-Witten map \cite{sw}.
In \refb{strength} the $\star$ product is the Moyal-Weyl product

\be
\label{moyal}{(f\star g)(x)=e^{{i\over 2}\theta^{ij}\partial_i{\partial'}_j}
f(x)g(x')|_{x=x'}}
\ee
The relation between this noncommutative field strength and the commutative
one we started from is encapsulated in the Born-Infeld action
$\sqrt{\det (g+ 2\pi\alpha' (B+F))}\,,$
the effective target space
action for an open string with slowly varying background fields
\cite{ft, acny}.
It turns out that 
\be
\label{bione}{\sqrt{\det (g+ 2\pi\alpha' (B+F))}\sim
\sqrt{ \det( G+ 2\pi\alpha'{\hat F})}}
\ee
where $g_{ij}$ is the `closed string metric' and $G_{ij}$
is the `open string metric' which appear in \refb{geeig}.

\section{Charged Open String}

We will now find the noncommutativity parameter
in the charged open string case
\cite{dn}.
Rather than studying the scaling limit of an open string
theory in a fully  non-abelian $U(N)$ background (which is still
difficult),
we will restrict ourselves to backgrounds that reside in the
$U(1)^N$ Cartan subgroup, but with different background $U(1)$ fields
on each brane. So we will discuss
the {\it charged} string, {\it i.e.} a string with different
$U(1)$ backgrounds at each end. We consider magnetic backgrounds with
constant field strength. In this case we have the exact mode expansion
to start with, originally derived in \cite{acny} and \cite{an}, and we
use it to
compute the  propagator on the disk. Starting from this propagator, we
repeat the argument in \cite{sch} to  compute two
noncommutativity parameters, one at each end of the string.
This emphasizes the interpretation that the noncommutativity of the
$D$-brane worldvolume in the presence of a background $B$-field along the
brane is really
a property of the endpoint of the string, rather than a feature of the
worldvolume itself. For $N$ $D$-branes, there are $N$ different
noncommutativity parameters, even when the $D$-branes are coincident.
Hence, the short distance behavior of the operator products
of tachyon vertex operators which are inserted on the boundaries
defines  star products with different noncommutativity parameters.
These enter into the computation of the scattering amplitudes of the
scaling limit theory.

The  worldsheet action for the ``charged'' string
with different magnetic fields at each end is

\ben
\label{wsacttwo}
S = {1\over{4\pi\alpha'}}\int_{\Sigma}
g_{\mu\nu}\partial_\alpha X^\mu \,\partial^\alpha X^\nu
-{i\over 2} \int_{-\infty}^{\infty} dt \,
(B^{(1)}_{ij} X^i \partial_t X^j\,|_{\sigma= 0}
+ \, B^{(2)}_{ij} X^i \partial_t X^j\,|_{\sigma= \pi} \,)\,.
\een
Here $0\le\mu,\nu\le 25\,$
and the open strings end on $D p$-branes in the $(0,i)$
directions for $1\le i\le p$.
Variation of \refb{wsacttwo} gives the equations of motion
for the worldsheet field
 \be
\label{eomx}{(\partial_\sigma^2 + \partial_t^2) \,X^\mu(\sigma, t) = 0}
\ee
together with the boundary conditions at each end of the string 
\ben
\label{bcx}
&& g_{ij} \partial_\sigma X^j +
2\pi i\alpha' B^{(1)}_{ij} \partial_t X^j \,|_{\sigma = 0} = 0\nonumber \\
&& g_{ij} \partial_\sigma X^j -
2\pi i\alpha' B^{(2)}_{ij} \partial_t X^j \,|_{\sigma = \pi} = 0
\een
and
\ben
\label{bco}
&&\partial_\sigma X^0 \,|_{\sigma =0,
\sigma = \pi} = 0\,\nonumber \\
&&X^I \,|_{\sigma =0} = 0\,,\quad p+1\le I\le 25\,.
\een

Note that the worldsheet action for the neutral string,
whose propagator in the directions along the $D p$-brane
is given in \refb{anprop},
is a special case of \refb{wsacttwo} with $B^{(1)}_{ij}=-B^{(2)}_{ij}$.
For simplicity, we
now specialize to the case where the ends of the string
live on  $D2$-branes,
{\it i.e.} $i=1,2$. In this case the magnetic fields have only one
component and we relabel them as
$B^{(1)}_{12} = q_1 B_{12}$ and $B^{(2)}_{12} = q_2 B_{12}$,
where $q_1 + q_2 \ne 0$.

In a basis given by
$X^\pm(\sigma, t) \equiv  X^1 (\sigma, t) \pm i  X^2 (\sigma, t)$,
the charged string normal mode expansion \cite{acny,an} can be written as
\be
\label{nmp}{ X^+ (z,\bar z) = x^+
+ {\textstyle {i\over 2}}\sqrt{2\alpha'}\sum_{r\in {\cal Z} + A}
{a_r\over r} ( z^{-r} + \bar z^{-r})
-{\textstyle {1\over 2}}\sqrt{2\alpha'} B \sum_{r\in {\cal Z} + A}
{a_r\over r} ( z^{-r} - \bar z^{-r} )}
\ee

\be
\label{nmm}{ X^- (z,\bar z) = x^-
+ {\textstyle {i\over 2}}\sqrt{2\alpha'}\sum_{s\in {\cal Z} - A}
{\tilde a_s\over s} ( z^{-s} + \bar z^{-s})
+{\textstyle {1\over 2}}\sqrt{2\alpha'} B \sum_{s\in {\cal Z} - A}
{\tilde a_s\over s} ( z^{-s} - \bar z^{-s} )\,.}
\ee
with  commutation relations  for the zero modes given by

\ben
\label{cscr}
&&[ a_r , \tilde a_s ]  =
2 G \,r \,\delta_{r,-s}\,;\qquad
[ a_r , a_{r'} ] = 0 = [ \tilde a_s , \tilde a_{s'} ]\,;\nonumber \\
\nonumber \\
&&[ x^+, x^- ] =  - {2 \over {(q_1 + q_2) B_{12}}}\,;\qquad
[ a_r , x^{\pm} ] = 0 = [ \tilde  a_s , x^{\pm} ]\,.
\een
Here $g_{ij} = g^{-1}\delta_{ij}$, $G_{ij} = G^{-1}\delta_{ij}$, where now
\be
\label{basta}
G = {g\over{1+B^2}}\,, \quad B = g q_1 2\pi\alpha' B_{12}\,,\quad  
A = {1\over \pi} ( \arctan B + \arctan {q_2\over q_1} B )\,.
\ee
Unlike the neutral string case, 
oscillators $a_r,\tilde a_s$ are non-integrally moded, since
$r = n+A$,  $s = n-A$, for $n\in {\cal Z}$.
We use the mode expansions computed above
to compute the charged propagator. The details of this calculation can be
found in \cite{dn}.
For $|z|> |\zeta|$, we find it is given by

\ben
\label{pmprop}
&& < X^+ (z,\bar z) X^- (\zeta,\bar\zeta) > \equiv
\langle \beta\, | X^+ (z,\bar z) X^- (\zeta,\bar\zeta) |\,\alpha\,
\rangle\nonumber \\
&&= {-2\alpha'\pi g\over {B + {q_2\over q_1} B}}
- 2\alpha' G {\textstyle{1\over A}} (\zeta^A + \bar \zeta^A - 1)
+ 2 i \alpha' G B {\textstyle{1\over A}} (\zeta^A - \bar \zeta^A)\nonumber \\
&& +\alpha' G \quad[ f({\textstyle{{\zeta\over z}}}) +
f({\textstyle{\bar\zeta\over {\bar z}}})
+ f({\textstyle{\zeta\over {\bar z}}})
+ f({\textstyle{\bar\zeta\over z}}) ]\nonumber \\
&& +\alpha' G \, B^2 \quad[ f({\textstyle{{\zeta\over z}}}) +
f({\textstyle{\bar\zeta\over {\bar z}}})
- f({\textstyle{\zeta\over {\bar z}}})
- f({\textstyle{\bar\zeta\over z}}) ]\nonumber \\
&& + 2 i\alpha' G \, B \quad
[ - f({\textstyle{\zeta\over {\bar z}}})
+ f({\textstyle{\bar\zeta\over z}}) ]
\een
where
\be
\label{defin}
{f(\rho) \equiv \sum_{r=n+A;\, n\ge 0} {\rho^r\over
r}\,;
\qquad\quad \lim_{A\rightarrow 0} f(\rho) = - \ln (1 - \rho)\ +
\lim_{A\rightarrow 0} {\rho^A\over A}\,.}
\ee

As a first step toward computing the commutator on the boundary, 
we need to compute the propagator on the boundary.
We distinguish the two boundary regions of the open string disk as follows.
On the boundary $\sigma = 0$, we have $z = |z| = \tau$ and
$\zeta = |\zeta | = \tau'$ so $\tau,\tau'>0$; while
on $\sigma = \pi$, then $z = |z| e^{i\pi}= \tau$ and
$\zeta = |\zeta |  e^{i\pi} = \tau'$ so here $\tau,\tau'<0$.

For $|z|> |\zeta|$, and on the boundary $\sigma = 0$, the propagator is
\ben
\label{bzero}
< X^+ (z,\bar z) X^- (\zeta,\bar\zeta) > |_{\sigma = 0}
&& =  {-2\alpha'\pi g\over {B + {q_2\over q_1} B}}\quad
+ 4\alpha' G \sum_{n=0}^\infty {1\over {n+A}}
({\textstyle{{\zeta\over z}}})^{n+A} \nonumber \\
&& +
{2\alpha' G \over A} - {4\alpha' G \over A}\,\zeta^A\,,
\een
\ben
< X^- (z,\bar z) X^+ (\zeta,\bar\zeta) > |_{\sigma = 0}
&&=  {2\alpha'\pi g\over {B + {q_2\over q_1} B}}\quad
+ 4\alpha' G \sum_{n=1}^\infty {1\over {n-A}}
({\textstyle{{\zeta\over z}}})^{n-A}\nonumber \\
&& + {2\alpha' G \over A} - {4\alpha' G \over A}
\, z^A\,.
\een
By using the above equations, we
compute the commutator  at the  $\sigma = 0$ end of the string.
\ben
\label{scho}
&&[ X^+ (\tau) ,  X^- (\tau)]
=T (  X^+ (\tau) \, X^- (\tau^-) -  X^+ (\tau)  X^- (\tau^+) )\nonumber \\
&&\equiv  \lim_{\epsilon\rightarrow 0}
(  < X^+ (\tau) \, X^- (\tau -\epsilon ) > -
< X^- (\tau + \epsilon )\, X^+ (\tau) > )
\, , \quad ({\rm for}\, \epsilon > 0)\nonumber \\
&&=  \lim_{\epsilon\rightarrow 0} \,
( \quad {-4\alpha'\pi g\over {B + {q_2\over q_1} B}}\quad
+  4\alpha' G [ \sum_{n=0}^\infty {1\over {n+A}}
({\textstyle{{\tau - \epsilon \over \tau }}})^{n+A}
\, -  \sum_{n=1}^\infty {1\over {n-A}}
({\textstyle{{\tau \over \tau + \epsilon }}})^{n-A} \,] \nonumber \\
&& - {4\alpha' G\over A} (\tau -\epsilon )^A
+ {4\alpha' G\over A} (\tau +\epsilon )^A \quad )\nonumber \\
&& =  {-4\alpha'\pi g\over {B + {q_2\over q_1} B}}\quad
+  4\alpha' G \,\pi \cot {\pi A} \nonumber \\
&& =   -4\alpha'\pi (g)^2 {q_1 2\pi\alpha' B_{12}\over
{1 +  (g)^2 (q_1 2\pi\alpha' B_{12})^2}}\nonumber \\
&& =  2 \Theta ^{12}\,.
\een
Notice that $\Theta^{12}$ is the same expression that appears
in the neutral string. The analogous calculation at $\sigma=\pi$
provides a different commutator.
\ben
\label{schpi}
&&[ X^+ (\tau) ,  X^- (\tau)]
 = T (  X^+ (\tau) \, X^- (\tau^-) -  X^+ (\tau)  X^- (\tau^+) )
\nonumber \\
&&\equiv  \lim_{\epsilon\rightarrow 0}
(  < X^+ (\tau) \, X^- (\tau +\epsilon ) > -
< X^- (\tau - \epsilon )\, X^+ (\tau) > )
\, , \quad ({\rm for}\, \epsilon > 0)\nonumber \\
&&= -4\alpha'\pi (g)^2 {q_2\pi\alpha' B_{12}\over
{1 +  (g)^2 (q_2 2\pi\alpha' B_{12})^2}}\nonumber \\
&&= 2 \tilde\Theta ^{12}\,.
\een

In the limit $q_1\rightarrow -q_2$, then
$ \tilde\Theta ^{12} \rightarrow - \Theta ^{12}$.
Indeed in the neutral string case,
where both ends of the string are on the same $D$-brane,
the noncommutativity parameter at one end of the string is
equal to minus that of the other end.
For $U(N)$ Chan-Paton factors, the background magnetic fields
can take on N possible values, giving rise to $N$ noncommutativity
parameters.

As in  \cite{sw}, one can show that in the scaling limit
 ($\alpha'\rightarrow 0$,
keeping $G$ and $\Theta^{12}$ fixed),  the operator product expansion
reduces to the star product \cite{dn}
\ben
\label{scaltach}
 e^{ip\cdot X(\tau)}   \,  e^{iq\cdot X(\tau')}
&\sim & e^{ -\Theta^{12} (p_- q_+  - p_+ q_- )} \,
: e^{i (p + q) \cdot X(\tau')} :\nonumber \\
&\equiv & e^{ip\cdot X(\tau')}   e^{iq\cdot X(\tau')}\,.
\een
For $\sigma = \pi$ the same equations
will hold with $\Theta^{12}$
replaced by $\tilde\Theta^{12}$.

The same results about noncommutativity parameters
can  be derived by starting with the charged string
propagator on the annulus, as calculated
in \cite{laidlaw} from a charged string annulus
propagator. Since the result is a short distance effect, it is
independent of the topology of the worldsheet.

\section{The Spectrum in the Scaling Limit and a Matrix Model}

Our explicit knowledge of the mode expansion
allows us to compute the spectrum of the gauge theory obtained
in the scaling limit, and compare it to one predicted by the matrix model.
In addition to the $U(1)^N$ massless gauge bosons, we find
charged vector states that survive for each Landau level.
The states of the limiting non-abelian
noncommutative gauge theory are no longer massless, but rather
tachyonic or massive.

As in \cite{sw}, we consider the scaling limit
$g^{-1}\rightarrow \epsilon $ and
$\alpha'\rightarrow \sqrt{\epsilon}$, for $\epsilon\rightarrow 0$.
Actually this limit means letting the dimensionless quantity
$\alpha' B_{12} \rightarrow \sqrt{\epsilon}$, while keeping $B_{12}$ fixed.
Then we have that the noncommutativity parameters are
finite in the scaling limit and are given by
$\Theta^{12}\rightarrow (- q_1 B_{12} )^{-1}$,
$\tilde \Theta^{12}\rightarrow ( - q_2 B_{12} )^{-1}$.

One can see from \refb{basta} that
$\tan {\pi A} = {{B + {q_2\over q_1} B}\over
{1 - {q_2\over q_1} B^2}}$. In the scaling limit,

\be
\label{tlim}\tan {\pi A} \rightarrow -{{(q_1+q_2) {\sqrt{\epsilon}}}\over
2\pi q_1 q_2 }\quad \, A\rightarrow -{{(q_1+q_2) {\sqrt{\epsilon}}}\over
2 \pi^2 q_1 q_2 }\,.
\ee
Using these formulae, one finds \cite{dn} that the mass formulae for the
two polarization states  of the
charged vectors are  finite in the scaling limit
and behave as

\ben
\label{mtach}
&&{\rm mass}^2
=-{1\over 2\alpha'} A (1+A) \rightarrow {1\over 2}{{(q_1+q_2) B_{12}}
\over 2 \pi^2 q_1 q_2 } =
- {(q_1+q_2) G
\over q_2 \Theta^{12}} \,,\nonumber \\
&&{\rm mass}^2
={1\over 2\alpha'} A (3-A))
\rightarrow -{3\over 2}{(q_1+q_2) B_{12}\over
2\pi^2 q_1^2 q_2 \Theta^{12}} =
3 {(q_1+q_2) G
\over q_2 \Theta^{12}}\,.
\een
Also in the spectrum are the charged vectors at different Landau levels. 
For each charged boson, the two polarizations have different masses.
They differ from those of \refb{mtach} by integer multiples
of ${2(q_1+q_2)G\over q_2 \Theta^{12}}$.
(Notice that each Landau level has infinite degeneracy).
All other states in the charged string spectrum become infinitely
heavy and decouple in the scaling limit.
So the complete spectrum of our $U(1)^N$ noncommutative field theory,
which is derived from $N$
neutral and $N^2 - N$ charged string sectors, is described
by $N$ massless neutral gluons and the charged vectors above.

We have seen that our theory, which is a $U(2)$ gauge theory expanded
around a $U(1)\times U(1)$ background, has a scaling limit.
The next question is  to find the corresponding action.

It turns out that
the noncommutative gauge theory action for
the $U(1)\times U(1)$ case can be described by a matrix model.
A matrix theory action in terms 
of two infinite-dimensional matrices $X^1, X^2$ is
$$ S = -{1\over {2 g_{YM}^2}}Tr_{\cal H} [X^1, X^2]^2 $$
where $X^1, X^2$ are some operators acting in a Hilbert space ${\cal H}$.
Its equations of motion are
$$ \delta_{k\ell}[X^k , [ X^\ell, X^j ] ] = 0\,.$$
A solution \cite{Nekrasov}
is $X_{\rm sol}^i$ given by a $2\times 2$ block matrix
of the form
$$X_{\rm sol}^i = \pmatrix{ y^i & 0 \cr 0 & z^i } $$
where $y, z$ satisfy the Heisenberg algebras
$$[y^i, y^j] = i {\theta}_1^{ij} {\bf 1}$$
$$[z^i, z^j] = i {\theta}_2^{ij} {\bf 1}\,,$$
{\it i.e.} $X^1_{\rm sol} + i X^2_{\rm sol} =
\pmatrix{ {\sqrt{2\theta_1}} a_1 & 0\cr
0 & {\sqrt{2\theta_2}} a_2}\,.$ Above, $a_1, a_2$ are the creation and annihilation operators in the Hilbert space $\cal H$.
Then it can be shown \cite{Nekrasov} by expanding
the equations of motions around
$ X^i = X^i_{\rm sol} + y^i $ and computing the coefficients
of the terms quadratic in $y^i$,
that the spectrum of fluctuations about this classical
solution coincides with the ${\alpha'} \to 0$ limit of the string
spectrum described above. One needs to 
identify
$$\theta_1 -\theta_2   =  {(q_1 + q_2) G \over q_2 \Theta^{12}},$$
where the RHS follows the notation from \refb{mtach} . 

\section{Time-dependent Background}

In most of the examples currently known,
the noncommutativity parameter is constant.
An obvious task is to look for time-dependent noncommutativity parameters.
In \cite{dn1} we studied an
open string model, whose target space has a plane wave metric
supported by a time-dependent Neveu-Schwarz two-form potential.
This background was studied in \cite{nw} for closed
strings, while in \cite{dn1} we looked at the open string version.
   We quantized the sigma-model in light-cone gauge,
computed the worldsheet propagator, and used it to derive
a time-dependent noncommutativity parameter. Indeed, for large values of
the time parameter, this model resembles
a neutral string in a constant background $B$ field,
hence it is a good candidate for spacetime noncommutativity.

We do not solve the exact open mode expansion in closed form, but
we compute it as a power series
in a suitable parameter $\mu$. This expansion is adequate to
show noncommutativity. To do that, we follow the same
strategy adopted for the neutral and charged string.
We compute a mode expansion, derive from it
a worldsheet propagator on the disk, and
evaluate the commutator on the boundaries to find
a time-dependent noncommutativity parameter.

The worldsheet action coupling a string to a general metric and
background Neveu-Schwarz field is

\be
\label{poly}{{S =
\int_\Sigma\, d\tau d\sigma
\,[\,{\sqrt{-\gamma}}\gamma^{\alpha\beta}\, G_{MN} \partial_\alpha
X^M \partial_\beta X^M + B_{MN} \epsilon^{\alpha\beta} \partial_\alpha X^M
\partial_\beta X^N \,]}}
\ee
where we choose the string worldsheet $\Sigma$ with Lorentz signature,
and have rescaled the scalar worldsheet fields by
$(2\sqrt{\pi\alpha'})^{-1}$
so that the $X^M$ are dimensionless.
We consider the time-dependent background provided by
the model based on a non-semi-simple group discussed in \cite{nw},
and adopt the same notation,
with $X^M = (a_1,a_2,u,v)$,
and $u$ being identified with the time in the target space.

The background field $G_{MN}$ and $B_{MN}$ are given by
 \be
\label{bmet}{G_{MN} =\pmatrix{1&0&{a_2\over 2}&0\cr
0&1&{\hskip-3pt-{a_1\over 2}}&0\cr
{a_2\over 2}&{\hskip-3pt-{a_1\over 2}}&b&1\cr
0&0&1&0\,\cr}\,,
\qquad B_{MN} =\pmatrix{0&u&0&0\cr
-u&0&0&0\cr
0&0&0&0\cr 0&0&0&0\cr}\,.}
\ee
The Lorentz signature target space metric
$G_{MN}$ can be recognized as a plane wave metric \cite{nw}.
The background is time-dependent because of the 
the $u$-dependence of $B_{12}$.
In \cite{nw} it was  shown  that  this model is exactly conformally
invariant ({\it i.e.} to all orders in $\alpha'$) by checking
the one-loop $\beta$ function equations for the closed string backgrounds,
and then proving that  there were no higher loop contributions.

Here, since we are interested in noncommutativity,
we consider open string boundary conditions.
This case is not conformally invariant.
\footnote{A conformally invariant version of
\refb{bmet} is studied in \cite{stt},
but its noncommutativity parameter although
non-constant, is not time-dependent.}
The background \refb{bmet} satisfies the
the Born-Infeld field equations for $N\ne u$,
\be
\label{bi}{(D_M F_{NL}) ( 1 - F^2)^{-1 \,LM} = 0}
\ee
where $ ( 1 - F^2 )^{-1\,LM} = ( 1 + F )^{-1\,LP}\, G_{PN}\,
(1 - F)^{-1\,NM}$ and
$( 1 - F)_{MN} \equiv G_{MN} - 2\pi\alpha' F_{MN}\,.$
In our case $F_{MN} = B_{MN}$.
Using  \refb{bmet}, one can check that
the nonvanishing components of the Ricci tensor and
affine connections are $R_{uu} = -\half$,
$ \Gamma^i_{uj} = \half\epsilon^i_{\,j}$, $\Gamma^v_{ui} = -{a^i\over 4}\,.$
It follows that
${(D_M F_{NL}) ( 1 - F^2)^{-1 \,LM}
= \epsilon_{ij}  ( 1 - F^2)^{-1 \,ju} = 0\,,}$ for $N\ne u$.
But
\be
\label{nbi}
{(D_M F_{uL}) ( 1 - F^2)^{-1 \,LM} = -{u\over{ 1 + (2\pi\alpha' u)^2}}
\,.}
\ee
In terms of the $a_i, u, v$ variables, the sigma model action is
\be
\label{polynw}S =
\int_\Sigma\, d\tau d\sigma
\,[\,{\sqrt{-\gamma}}\gamma^{\alpha\beta}\,(\partial_\alpha
a^i\partial_\beta
a^i + 2 \partial_\alpha u\partial_\beta v +
b \partial_\alpha u\partial_\beta u 
+\epsilon_{ij}\partial_\alpha u\partial_\beta a^i a^j )
+\epsilon^{\alpha\beta} \epsilon_{ij} u
\partial_\alpha a^i \partial_\beta a^j\,] .
\ee

Although our background is not conformally invariant,
we will consider a light-cone version of the sigma model
in order to study open string propagators in a B-field
with linear time dependence.
We identify the target space time $u$ with the
worldsheet time $\tau$ via $u=\mu\tau$, where $\mu$ is a dimensionless
parameter.
The equations of motion and boundary conditions for the
transverse fields $a^i$ written in terms of $X \equiv a^1 + i a^2$ and
$\tilde X \equiv a^1 - i a^2$ become:

\ben
\label{eomlg}
&& \Box X - i \mu ( \partial_\sigma X - \partial_\tau  X )
= 0\,,\qquad
\Box \tilde X +
i \mu ( \partial_\sigma \tilde X - \partial_\tau  \tilde X ) = 0\,,\nonumber \\
&&[\, \partial_\sigma X + i\mu\tau \partial_\tau X \,]\,|_{\sigma=0,\pi}
 =  0\,,\qquad
[\,\partial_\sigma \tilde X - i\mu\tau \partial_\tau \tilde X\,]
\,|_{\sigma=0,\pi}
= 0\,,
\een
where $\Box\equiv -\partial_\tau^2 + \partial_\sigma^2 =
4 z \bar z \partial_z\partial_{\bar z}\,.$

For large $\tau $ (so that $\tau$ can be considered constant),
notice the similarity of the boundary condition in \refb{eomlg}
with the boundary condition for an open string in a background
$B$ field.
Since in the latter case the noncommutativity parameter is proportional to
the background, this suggests we should expect here
a noncommutativity parameter which depends on time.

The solution of \refb{eomlg} is given by the normal mode expansion
for the transverse coordinates $X$ and $\tilde X$,
to first order in $\mu$:
\ben
\label{nmex}
&&X(\sigma,\tau)
=x_0 + a_0 [ \tau + \mu ( - i\tau\sigma
+ {i\over 2} \tau^2) ]\nonumber \\
&& + \sum_{n\ne0} a_n e^{-in\tau}
[ {i\over n} \cos n\sigma +  \mu ( ( -{1\over 2n^2} - i {\tau\over n} )
\sin n\sigma + ( {i\over 2n^2} + {(\sigma - \tau) \over 2n} ) \cos n\sigma )
] \\
&& \tilde X(\sigma,\tau)
=  \tilde x_0 + \tilde a_0 [ \tau - \mu ( - i\tau\sigma
+ {i\over 2} \tau^2) ]\nonumber \\
&& + \sum_{n\ne0} \tilde a_n e^{-in\tau}
[ {i\over n} \cos n\sigma - \mu ( ( -{1\over 2n^2} - i {\tau\over n} )
\sin n\sigma + ( {i\over 2n^2} + {(\sigma -\tau) \over 2n} )
\cos n\sigma )]\,.\,
\een

\section{Time-dependent Noncommutativity}

Having found a mode expansion,  the propagator on the disk can be computed
along the lines of  {\cite{dn}}. We will
use the notations  $z=e^{i(\tau + \sigma)}$, $\bar z=e^{i(\tau - \sigma)}$,
$\zeta=e^{i(\tau' + \sigma')}$ and $\bar\zeta=e^{i(\tau' - \sigma')}$
Then for $|z|> |\zeta|$, the propagator to order $\mu$ is
\eject
\ben
\label{propa}
< X (z,\bar z) \tilde X (\zeta,\bar\zeta) >
& = & - i 4\alpha' \,
(\tau + \mu ( -i \tau\sigma + {i\over 2}\tau^2) )\nonumber \\
& + & 4\alpha' \sum_{n=1}^\infty  e^{-i n (\tau - \tau')}
[\,{1\over n} \cos n\sigma \cos n\sigma'\nonumber \\
& + & i\mu \cos n\sigma' (\, ( {1\over 2n^2}
+ {i\tau\over n} ) \sin n\sigma
- ( {i\over 2n^2} + {(\sigma - \tau)\over 2n} )\cos n\sigma \,)\nonumber \\
& - & i\mu \cos n\sigma (\, ( {1\over 2n^2}
- {i\tau'\over n} ) \sin n\sigma'
+ ( {i\over 2n^2} - {(\sigma' - \tau')\over 2n}) \cos n\sigma' \,) \nonumber \\
& - & {\mu\over n^2} \cos n\sigma \cos n\sigma'\,]\nonumber \\
& + & \mu ( c_1\tau + c_0 )\,.
\een

We are free to add the function $ \mu (c_1 \tau + c_0 ) $ to the expression
since it does not affect the equation of motion or the boundary
condition for the propagator to first order in $\mu$.
For  $|z|> |\zeta|$, the expression for
$< \tilde X (z,\bar z) X (\zeta,\bar\zeta) >$ is given by
letting $\mu\rightarrow -\mu$ in the above propagator.
In the $\mu\rightarrow 0$ limit, these propagators reduce to the
open bosonic string propagator
$\lim_{\mu\rightarrow 0} < X (z,\bar z) \tilde X (\zeta,\bar\zeta) >
= -2\alpha' ( \,\ln |z-\zeta| + \ln |z-\bar\zeta| \,)$.

We evaluate the propagator  on the worldsheet boundary
at $\sigma = 0$ and $\sigma = \pi$. We denote the points on the boundary
at $\sigma =0$ with $z =  e^{i\tau}
\equiv {\cal T}$,
and $\zeta = e^{i\tau'} = {\cal T'}$.
We get on the $\sigma =0$ boundary
\ben
\label{probn}
< X (z,\bar z) \tilde X (\zeta,\bar\zeta) >|_{\sigma = 0}
& =& -i4\alpha'\, ( \tau + \mu {i\over 2}\tau^2)  +
\mu ( c_1\tau + c_0)\nonumber \\
& - & 4\alpha' \ln ( 1 - e^{- i (\tau - \tau')} )
- 2\alpha' \mu
\,i (\tau - \tau')\, \ln ( 1 - e^{-i (\tau - \tau')})\nonumber \\
& = & - 4 \alpha' \ln ({\cal T} - {\cal T'})\nonumber \\
& + & \mu ( - 2 \alpha' \ln^2 {\cal T}
- 2 \alpha'
\ln ({{\cal T}\over {\cal T'}}) \, \ln ( 1 - {{\cal T'}\over {\cal T}})
\, + ( -c_1 i \ln {\cal T} + c_0) )\nonumber \\
< \tilde X (z,\bar z) X (\zeta,\bar\zeta) >|_{\sigma = 0}
& = & -i4\alpha'\, ( \tau - \mu {i\over 2}\tau^2) -
\mu ( c_1\tau + c_0)\nonumber \\
& - & 4\alpha' \ln ( 1 - e^{- i (\tau - \tau')} )
+ 2\alpha' \mu
\,i (\tau - \tau') \, \ln ( 1 - e^{-i (\tau - \tau')})
\een
Then at $\sigma = 0$ the commutator is 

\ben
\label{scho}
[ X ({\cal T}) ,  \tilde X ({\cal T})]
& = & T (  X ({\cal T}) \, \tilde X ({\cal T}^-) -  X ({\cal T})
\tilde X ({\cal T}^+) )\nonumber \\
&\equiv & \lim_{\epsilon\rightarrow 0}
(  < X ({\cal T}) \, \tilde X ({\cal T}-\epsilon ) > -
< \tilde X ({\cal T} + \epsilon )\, X ({\cal T}) > )
\, , \quad ({\rm for}\, \epsilon > 0)\nonumber \\
& = & \mu \,
(-4 i \alpha' ) ( \pi \ln {\cal T} - i\ln^2{\cal T})
\nonumber \\
& = & \mu \,  4\alpha' \,( \pi \tau +  \tau^2)
\equiv \Theta\,,
\een
where we chose $c_1 = 2\pi\alpha'$, $c_0 = 0$,
and use $\lim_{\epsilon\rightarrow 0} (\ln ( 1 + \epsilon)
\ln \epsilon ) = 0$.
The noncommutativity parameter $\Theta$ is time-dependent.

Similarly, at  $\sigma = \pi$ the commutator is

\ben
\label{schpi}
[ X ({\cal T}) ,  \tilde X ({\cal T})]
& = & T (  X ({\cal T}) \, \tilde X ({\cal T}^-) -  X ({\cal T})
\tilde X ({\cal T}^+) )\nonumber \\
&\equiv & \lim_{\epsilon\rightarrow 0}
(  < X ({\cal T}) \, \tilde X ({\cal T}+\epsilon ) > -
< \tilde X ({\cal T} - \epsilon )\, X ({\cal T}) > )
\, , \quad ({\rm for}\, \epsilon > 0)\nonumber \\
& = & (-i 4\alpha')\, \mu [\, -\pi \ln{\cal T}
- i \ln^2{\cal T} \,]\nonumber \\
& = & \mu \,\, 4\alpha' \, (\,-\pi \tau + \tau^2)
\een
Thus for small $\mu$, we have

\ben
\label{ncpmu}
\Theta & = &  \mu \,\, 4\alpha' \,( \pi \tau +  \tau^2)
\,\qquad {\rm at}\, \sigma = 0\,,\nonumber \\
\Theta & = & \mu \,\, 4\alpha' \, (\,-\pi \tau + \tau^2)\,
\quad {\rm at}\, \sigma = \pi\,.
\een
For small $\tau$, the theta parameter at the
$\sigma = 0$ end of the string is minus that at the $\sigma = \pi$ end.
This is the case for the neutral string in a constant background
$B$ field as well. In fact, although we have worked only to
lowest order in $\mu$, we can see directly from the equations of
motion and boundary conditions (in $z$, $\bar z$) variables
in the limit of large $z$, {\it i.e.}
large $i\tau$, a limit for which $z^{-1}\rightarrow\ 0$, that
the system reduces to the neutral string with
the identification $- \mu \tau = B$, a constant.
(In the large $\tau$ limit, we note that $\ln |z|$ is approximately
constant, in the sense that it is changing slowly, {\it i.e.} its derivative
$ |z|^{-1}$ is small. Therefore, for large $\tau$ the noncommutativity
parameter becomes constant, and our model is similar to the
neutral string.)
For large $\tau$, using the neutral string expressions,
we find the noncommutativity parameter
be time-dependent

\ben
\label{ncp}
\Theta & = & -4\alpha' \pi B = 4\alpha' \mu \pi \tau
\,\,\,\quad {\rm at} \,\,\,\sigma = 0\,,\nonumber \\
\Theta & = & 4\alpha' \pi B = - 4\alpha' \mu \pi \tau
\,\,\,\quad{\rm at} \,\,\sigma = \pi\,.
\een

We have shown that our model
exhibits noncommutativity for both small and large $\tau$.
The expectation is that the model will remain noncommutative
with a time-dependent noncommutativity parameter for all times.

\section{Conclusions}

The noncommutativity properties of string theories are interesting at many
levels.  In string theory, noncommutativity has lead to new insights and new
techniques.
From a mathematical point of view, noncommutativity in string theory is
interesting because of its connection to noncommutative geometries and
noncommutative algebras. Many open questions remain.
One is what is
the generalization of the Moyal-Weyl product in the case of multiple non
commutativity parameters, as in the case of non-abelian gauge theories.
What is
the generalization of the
map in \refb{map}? Partial answers have been attempted in
\cite{dy,haya,mukhi}, but much more needs to be understood. In fact, we might
be just at the beginning of our understanding of the relation between
strings and noncommutativity.
We suspect that the inclusion of additional terms in the worldsheet
action, such as mass terms or Ramond backgrounds \cite{van},
will lead to novel types of noncommutativities.
Probably there is a sort of
``noncommutative democracy'' among the various pieces
of the worldsheet action. This set of noncommutativity parameters,
one for each of the additional terms in the worldsheet action,
could be reflected in the interpretation of the open string field theory
star product as a continuous tensor product of Moyal products\cite{doug}.

Finally,  the most interesting aspect is that spacetime
noncommutativity discussed above offers
an insight on the structure of spacetime, as it implies
a spacetime uncertainty principle \cite{yonega}
$$\Delta X^\mu\Delta X^\nu\ge \alpha'\,.$$

\noindent
This relation suggests the existence of a spacetime cutoff at the Planck
scale, and hence possible deviations from the smoothness of spacetime at
small distances. It might well be that these aspects of Planck scale physics
lead to distinctive signatures observable in cosmology. In particular,
they could leave their mark on the spectrum of density fluctuations in the
early universe, and modify
the inflationary perturbation spectrum.

\end{document}